\documentclass[11pt,twoside]{article}
\usepackage{asp2004}
\usepackage{psfig}
\usepackage{epsf}
\usepackage{graphics}
\usepackage{lscape}
\begin{document}

\hyphenation{Dopp-ler}
\title{Simultaneous Doppler maps  of IP Peg in outburst}
\author{C. Papadaki$^{1,2}$, H.M.J. Boffin$^3$ \& D. Steeghs$^4$}
\affil{
$^1$Royal Observatory of Belgium, 3 av. Circulaire, 1180 Brussels, Belgium
$^2$Vrije Universiteit Brussel (OBSS/WE), Pleinlaan 2, 1050 Brussel, Belgium
$^3$European Southern Observatory, Karl-Schwarzschild-Str. 2, D-85748 Garching bei Munchen, Germany
$^4$Harvard-Smithsonian Center for Astrophysics, 60 Garden Street
Cambridge, MA 02138, USA}

\begin{abstract}
IP Pegasi is an eclipsing dwarf nova lying above the period gap with
an orbital period of 3.8h. It is the first cataclysmic variable to
show evidence of spiral arms in its accretion disc. We present new time-resolved echelle spectroscopy observations of IP Peg, covering the 3900-7700{\AA} range. This allows us to construct simultaneous Doppler Maps in 9 emission lines.
\end{abstract}

\section*{Observations}
Using EMMI on ESO's NTT, we obtained phase-resolved echelle spectroscopy of IP Peg in August 1999, when the system was in outburst.
Our series of 39 spectra, each with an
integration time of 300s, extends for more than 5.4 hours, i.e. 1.5 orbital
period. The large wavelength coverage, approximately from 3900 to
7700{\AA}, provides us with several emission lines whose simultaneous
study allows us to probe in detail the structure of the accretion disc
and the contribution of the secondary. 

The spectra were reduced using the echelle package routines in IRAF,
while the extraction of the orders and the wavelength calibration were
performed by the echelle package reduction task doecslit. The
continuum contribution was subtracted by fitting a cubic spline to the
appropriate parts of each order in each spectrum.

\vskip -0.2cm
\begin{figure}[!ht]
\plotone {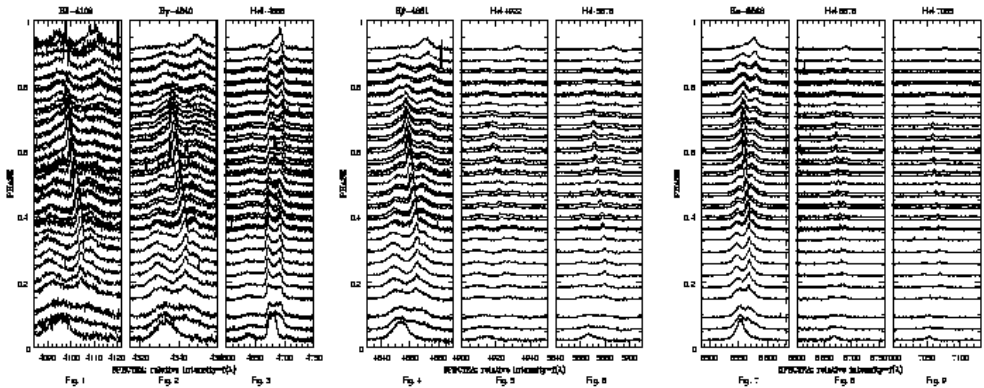}
\vskip -0.35cm
\caption {Phase folded spectra of the 9 most prominent emission
  lines}
\vskip -0.45cm
\end{figure}

\section*{Analysis}

IP Peg in contrast to other dwarf novae does not develop absorption
lines during outburst (Piche \& Szkody, 1989). A probable
explanation could be its high
inclination i$\approx$81$\deg$ that reduces the contribution of the accretion disc
absorption lines. From the output spectra no absorption lines were
identified, while the 9 most prominent emission lines were selected
for further studying. These include the Balmer series H$\alpha$,
H$\beta$, H$\gamma$, H$\delta$,
four HeI lines with wavelengths 7065, 6678, 5876, 4922{\AA} and one HeII
line at 4686{\AA}. The phase folded spectra of all the selected emission
lines are plotted in figure 1. The orbital ephemeris for the
calculation of the corresponding phases was taken from
Wolf et al. (1993).
\par
In figure 1, we detect the characteristic S-wave of a
low-velocity emission from the secondary, which reaches maximum at
phase 0.5 when the secondary is seen face-on. We also see a clear
variation in the line profiles and in particular a variation of the
separation between the red and blue-shifted emission peaks originating
in the accretion disc. 

\begin{figure}[!ht]
\plotone {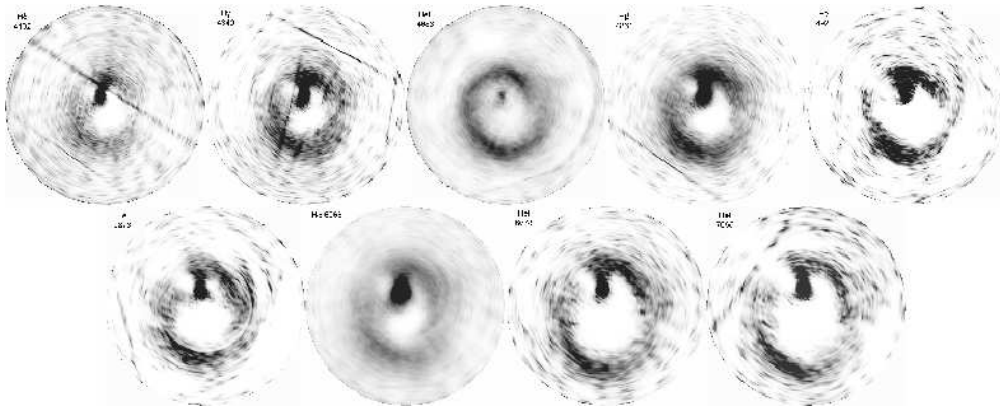}
\caption {Doppler maps of the corresponding lines of figure 1}
\vskip -0.165cm
\end{figure}

As a next step we made a preliminary attempt to apply Doppler
imaging on our unique set of spectra. Using Henk Spruit's code
(Spruit 1998, http://www.mpa-garching.mpg.de/\~{}henk/pub/dopmap),we
constructed Doppler maps (figure 2).  Despite some clear artifacts, which are, among others,
caused by prominent spikes near the emission lines (as seen in some
cases in figure 1) the two spiral arms are visible in each of
the Doppler maps. Depending on the emission line the intensity of the
spiral arms as well as of the secondary star varies. The comparison
between these different simultaneous maps will bring invaluable
information on the source of the spiral arms and the structure of the
accretion flow. This work is now in progress.

\end {document}